\documentclass[aps,pra,twocolumn]{revtex4-2}
\usepackage{graphicx}
\usepackage{bm}
\usepackage{amsmath}
\usepackage{esvect}
\usepackage[capitalize]{cleveref}

\providecommand{\abs}[1]{\left\lvert#1\right\rvert}

\providecommand{\bra}[1]{\langle #1 \rvert}
\providecommand{\ket}[1]{\lvert #1 \rangle}

\providecommand{\be}{\begin{equation}}
\providecommand{\ee}{\end{equation}}
\providecommand{\ba}{\begin{eqnarray}}
\providecommand{\ea}{\end{eqnarray}}

\usepackage{mathbbol}

\usepackage{tikz}
\usepackage{amsfonts,amssymb}
\usepackage{dsfont}
\usepackage{physics}
\usepackage{tocvsec2}

\newcommand{\beq}{\begin{equation}}
\newcommand{\eeq}{\end{equation}}
\providecommand{\dt}[1]{\frac{d#1}{dt}}

\usepackage{appendix}

\begin{document}

\title{Reconstructing the full modal structure of photonic states by stimulated emission tomography in the low-gain regime}
 
\author{ A. Keller$^{1,2}$, A. Z. Khoury$^3$, N. Fabre$^4$, M. Amanti$^1$, F. Baboux$^1$, S. Ducci$^1$ and P. Milman$^1$ }

\affiliation{$^{1}$Université Paris Cité, CNRS, Laboratoire Matériaux et Phénomènes Quantiques, 75013 Paris, France}
\affiliation{$^{2}$ Department de Physique, Université Paris-Saclay, 91405 Orsay Cedex, France}
\affiliation{$^{3}$ Instituto de F\'isica, Universidade Federal Fluminense, Niteroi, RJ 24210-346, Brazil  }
\affiliation{$^{4}$Departamento de Óptica, Facultad de Física, Universidad Complutense, 28040 Madrid, Spain}

\begin{abstract}
Stimulated emission tomography (SET) is a powerful and successful technique to both improve the resolution and experimentally simplify the task of determining the modal properties of biphotons. In the present manuscript we provide a theoretical description of SET valid for any quadratic coupling regime between a non-linear medium and pump fields generating photons by pairs in the low-gain regime. We use our results to obtain not only information about the associated modal function modulus but also its phase, for any mode, and we discuss the specific case of time-frequency variables as well as the quantities and limitations involved in the measurement resolution.

\end{abstract}
\vskip2pc 
 
\maketitle
\section{Introduction}

Photons are elementary excitations of the electromagnetic field in a given mode, which can combine polarization, frequency, the transverse position or momentum, among other degrees of freedom. As such, photons cannot be separated from the properties defined by the modes they occupy. These are actually good news: since modes can be decomposed in orthonormal basis, they can be manipulated and engineered using linear and non-linear devices, and different modes can be measured independently. Consequently, modes can be used not only to distinguish, isolate and manipulate quantum properties of the electromagnetic field but also to measure its statistics \cite{TrepsModes, Laurat}. 

The interplay between the electromagnetic field states and modes - or, more poetically, between light and color, if one refers to frequency modes - presents thus many facets, and modes are valuable tools to exploit different properties and aspects of the electromagnetic field. One example are non-locality tests using single photons, which rely on the independent measurement of the polarization modes of the quantum state of a photon pair in spatially separated modes. The correlations between polarization measurements of photons belonging to the same pair led to the demonstration of non-classical effects that are particular to quantum states of the field \cite{aspect_experimental_1982}.

 Polarization, as well as all modal attributes, are properties of the electromagnetic field which apply to both classical and quantum descriptions. The modal measurement statistics inherit the photon's quantum properties and can be used as a way to reveal different surprising aspects of quantum physics as, in this case, non-locality. In addition, mode manipulation and engineering can lead, through the same principles, to the modification of the photon statistics in particular measurements, controlling for instance the distinguishability between photons and tailoring effective commutation relations \cite{francesconi_engineering_2020, MultimodeSteve,PhysRevLett.105.200503}.

For photon pairs generated by parametric processes, as spontaneous parametric down conversion (SPDC) or four-wave mixing (SFWM), the modal structure of the photon pair is crucial to infer some quantum properties of the produced state, as entanglement. The conservation laws involved in the photon pair production lead to a correlated joint modal function and the number of independent orthogonal modes which are necessary to describe the pair's mode structure is directly related to the degree of entanglement of the state \cite{lamata_dealing_2005}. Consequently, experimentally inferring entanglement and using it as a resource demand a perfect knowledge of the joint modal structure of the photon pair \cite{donohue_quantum-limited_2018}. 

Finally, the manipulation of the mode basis of squeezed states can transform entangled states into separable ones and vice-versa, and this mode basis transformation enables multi-mode entanglement detection in continuous variables and their applications as quantum teleportation \cite{PhysRevA.49.1473}. Modes also play a key role in quantum state measurement and metrology \cite{PRXQuantum2, donohue_quantum-limited_2018}, since usual techniques as the homodyne detection, require a near perfect mode overlap between the probed state and the reference one in order to be reliable \cite{Gil-Lopez:21, PhysRevLett.109.053602, roslund_wavelength-multiplexed_2014}.  

The mode structure of a quantum field depends on the conditions of its generation, as the properties of the medium producing it and the complex amplitude of auxiliary fields. One of the simplest configuration leading to the generation of non-classical states of light consists of non-linear interactions that can be described by non-degenerate quadratic operators with a complex mode structure : 

\begin{equation}\label{H}
\hat S = \gamma \int \int L({\bf k},{\bf k'})\hat a_s^{\dagger}({\bf k})\hat a_i^{\dagger}({\bf k'}){\rm d}{\bf k}{\rm d}{\bf k'}. 
\end{equation} 

In the physical process associated to (\ref{H}), photons are created in pairs and distributed in different auxiliary modes (as polarization or propagation direction) which are labeled $s$ and $i$, for signal and idler, with $\int \int |L({\bf k}, {\bf k'})|^2 {\rm d}{\bf k} {\rm d}{\bf k'}=1$. We chose to discuss the non-degenerate case for generality, but our results also apply to the degenerate one. The modal variables ${\bf k}$ and ${\bf k'}$ can refer to any mode, as frequency, time, propagation direction, among others. Also, the integral over modes ${\bf k}$ and ${\bf k'}$ for the signal and the idler fields may as well involve a sum over discrete modes, as the orbital angular momentum or polarization. We used only the integral form to simplify the notation. 

In Eq. (\ref{H}), $\gamma$ is a coupling constant, given by the product of two physically independent quantities, that we call ${\cal A}$ and ${\cal \chi}$,  so that we can write $\gamma = {\cal A}\chi$. While ${\cal A}$ depends on physical parameters that can be controlled at each run of an experiment, as the complex amplitude of auxiliary fields, $\chi$ relates to the gain, and is considered as a constant in time that depends only the non-linear device used in the experiment, as the material's size and geometry. The distinction between the two parameters is important since in the present manuscript, we'll always consider that $|\chi|^2 \ll 1$ (see Appendix A), while $\gamma$ is arbitrary.

Eq.(\ref{H}) can describe different physical processes, such as four-wave mixing, or a general parametric process, for instance. In both of the mentioned cases, Eq.(\ref{H}) is an approximation of the non-linear interaction describing the conversion of photons from classical fields (coherent states), called pump(s), into two photons, mediated by a material medium. The complete derivation of Eq. (\ref{H}) can be found in Appendix A, and involves considering that the pump's state remains a coherent state (continuous-wave or pulsed) and can be treated as classical during its interaction with the medium. This assumption is important since, in this case, Eq. \eqref{H} leads to an output quantum state $\ket{\psi_{\text{out}}}$ that can be written, in the low-gain regime, as (see appendix~A)

\begin{equation}
\ket{\psi_{\text{out}}} = e^{ \hat{S}-\text{h.c.}}\ket{\psi_{\text{in}}} = \hat U\ket{\psi_{\text{in}}}, 
\label{U}
\end{equation} 
where $\ket{\psi_{\text{in}}}$ is the total state of the electromagnetic field before the non-linear interaction. 

If $\gamma \ll 1$ (which is the case, for instance in SPDC), we can consider only the first terms of the expansion of Eq. (\ref{U}) in powers of $\gamma$; in this regime, the creation of photon pairs can be observed by post-selection using coincidence detection. Alternatively, but still in the low-gain regime, cavity effects can enhance the creation of squeezed states, that can be detected using homodyne detection, typically. These two regimes are contemplated by the present manuscript, that has as its main goal the measurement of the modal function $L({\bf k},{\bf k'})$. This function is determined by the medium and by the (possible) auxiliary fields' spectral properties, but doesn't depend on the auxiliary or on the pump field's amplitude.

So, from Eq. (\ref{U}), we see that if $\ket{\psi_{\text{in}}}$ is Gaussian, $\ket{\psi_{\text{out}}}$ will also be Gaussian. If the low-gain hypothesis no longer holds, time-ordering effects must be considered to obtain the evolution operator associated to the non-linear interaction, and this regime was extensively discussed in \cite{QuesadaTime, Quesada1, QuesadaPRX, BoydSqueezedBright}.

In the present manuscript, we provide general methods to infer full or partial information about the modal function $L({\bf k},{\bf k'})$ for any mode ${\bf k},{\bf k'}$, and we discuss in detail the case of the frequency properties of the field. As mentioned, the modal function is an essential ingredient for all the applications based on the non-linear interaction (\ref{H}). In particular, it determines how the produced states can be used as a resource for some quantum information, computation, communication and metrology protocols.

Given the interest of the task, several experimental methods have been proposed so far according to the specific type of mode one wants to access as well as the coupling strength regime. A good overview of the existing techniques can be found in \cite{ReviewKwiat}. 

\begin{figure}
 \begin{center}
\includegraphics[width=0.5\textwidth]{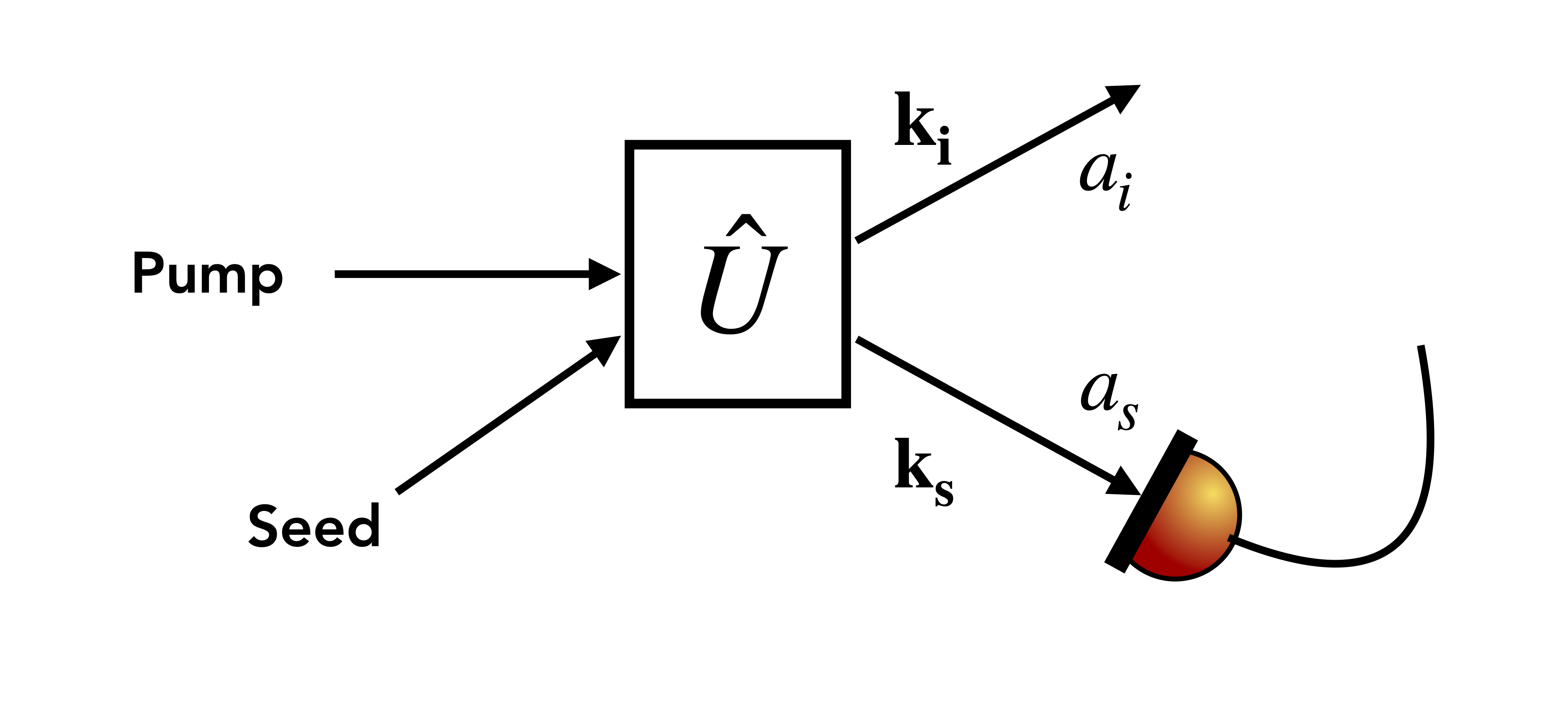}
\caption{\label{superpos}Schematics of the Stimulated Emission Tomography (SET).}
\end{center}
\end{figure}

Among these, an interesting approach is stimulated emission tomography (SET). Introduced in \cite{Sipe}, SET was used with great success to infer mode properties of photon pairs using relatively simple experimental techniques avoiding coincidence counts and low intensity signals - and consequently, improving the measurement resolution and reducing the integration time. SET can be applied to all types of modes, and its main principles are sketched in Fig.~\ref{superpos}: a pump beam is sent through a non-linear medium, together with a seed beam in mode $i$. In its original formulation, the seed beam is prepared in a coherent state in a mode defined by ${\bf k'_i}$ in the support of the idler's mode with resolution ${\bf \delta k'}$, where ${\bf \delta k'}$ is much narrower that the idler mode's width. Then,  the intensity of the signal mode is detected with a resolution $\delta {\bf k_s}$ around a given value ${\bf k_s}$. The main result of \cite{Sipe} is to show that the obtained intensity is proportional to the coincidence detection of photon pairs that are produced in the spontaneous regime, {\it i.e.}, in the absence of a seed beam. The proportionality factor is given by the intensity of the seed beam, and this is the reason for the signal to noise improvement. Also, since the intensity of the seed beam is known,  one can easily recover from this measurement the absolute value of the modal function of the produced two-mode state, $|L({\bf k_s}, {\bf k'_i})|^2$ at points ${\bf k_s}$, ${\bf k'_i}$ defined by the seed mode choice and the signal beam mode selective detection.

Using SET to obtain full phase and amplitude information about the modal function associated to the signal and idler beams requires different experimental resources according to the mode one is interested in. SET can be straightforwardly used to completely characterize the polarization state generated in a non-linear process \cite{PolarSET}, since by changing the polarization of the seed beam and the one of the detected signal field, one can reconstruct the full two-photon polarization state with intensity measurements only. This idea can be extended to other degrees of freedom, as propagation direction, by discretizing them, as was done in \cite{BeyondSET}.  

As for the field's spectral properties, the situation is trickier. It is experimentally challenging to manipulate and shape the seed so as to provide information about the phase of the spectral function with intensity measurements only. A clever solution for this was provided in \cite{Jizan}, where the authors use a same classical beam as pump, seed and as a phase reference pulse. The original field is decomposed, multiplexed and chirped, and the field generated through the non-linear interaction is made to interfere with the reference beam with a phase difference that can be experimentally adjusted. The detected signal has the same form as one of the outputs of an interferometer, so by combining output signals with different phase  choices, one reaches a result which is proportional to the photon pair's Joint Spectral Amplitude (JSA), $L({\bf k},{\bf k'})= L(\omega,\omega')$. Alternative techniques to obtain phase information \cite{PhysRevLett.127.163601} and complement the SET measurements in its original formulation were also proposed and implemented in \cite{Beduini, Kyoon,Borghi, WalmsleySpectral}, but they require using coincidence measurements as well.

Our work is a general way to extend the SET technique so as it can provide, for arbitrary modes, full information about the produced field's modal function without the need to perform coincidence detection. This means that we show it is possible to obtain, not only information about the absolute value of the modal function, as in \cite{Sipe}, but also its phase, for any mode.  In order to do so, we present a slightly different theoretical model for the output field produced by the non-linear interaction than the one used for instance in \cite{Sipe, Mexican}. Furthermore, we consider the exact output state for an arbitrary effective coupling regime (arbitrary value of $\gamma$ in Eq. (\ref{H})). Finally, we combine the stimulated output field in both signal and idler modes using an interferometric configuration and compute the difference of intensities of the combined modes. The reasons for these modifications are twofold: first of all, since SET relies on classical intensity measurements, all the involved operations, interactions and elements are Gaussian. Thus, using in any step of the model non-Gaussian states may be confusing and inaccurate. The output state in the SET configuration is Gaussian, and must be considered as such, in spite of the fact that the spontaneous regime combined with post-selection by coincidence detection produces a non-Gaussian state, as is usually the case in SPDC. Secondly, keeping the full evolution operator in Eq. (\ref{U}) enables extending SET to other experimental situations where multi-mode squeezed states can be produced and measured, as for instance Optical Parametric Oscillators and homodyne detection.  We'll see how these modifications change the results obtained in \cite{Sipe}, which can be of course perfectly recovered by making a expansion on parameters that, surprisingly, depend not only on the coupling strength but also on the frequency correlation properties of the output field.

 We will thus present in the following the exact calculation of the signal intensity detection in the stimulated regime, using methods similar to the ones of \cite{SipeVelho, Mexican, Sipe}. In addition, we'll use our results to propose a method to interferometrically measure either the full modal function, including phase information, or its Fourier transform, according to the choice made on the seed's spectral properties.  We'll see that the conditions on the seed's spectrum so as to unveil the modal function or its Fourier transform are not exactly the same as in \cite{Jizan, Sipe} nor are simply the Fourier analogous to them. They are rather the analog of a two-party spectral homodyne detection. 

The manuscript is organized as follows: Section \ref{Direct} is devoted to the direct detection of the signal field, in the same spirit as the original formulation of SET. This section is important to fix notations and establish the main differences between our approach and the one of \cite{Sipe}. In  Section \ref{Inter} we use the obtained results to develop the principles of the interferometric detection. Finally, in Section \ref{JTASec}, we discuss the particular case of frequency and time measurements.

\section{Direct detection}\label{Direct}

The first measurement configuration we study is the same as in the original paper \cite{Sipe}, but with a slightly different approach. First of all, we consider that the non-linear interaction between the medium and the pump-beam can be described by the unitary operator appearing in (\ref{U}). 

The unitary operator is thus applied to the system's initial state $ \ket{\psi_{\text {in}}}$,  given by the vacuum state displaced by the seed, which is a coherent state:  $\ket{\psi_o}=\hat D(\alpha)\ket{0}$, where $\alpha$ is the complex amplitude of a coherent state in the seed's mode (See Appendix A for a derivation). The seed mode  $\alpha$ can also be represented as a superposition of modes. More specifically,  the associated multimode displacement operator is given by 

\begin{equation}\label{D}
\hat D(\alpha)= e^{\int ( \hat a^{\dagger}({\bf k})\alpha({\bf k}) - \hat a ({\bf k})\alpha^*({\bf k})){\rm d}{\bf k}},
\end{equation}
where $|\alpha|^2=\int |\alpha({\bf k})|^2{\rm d}{\bf k} $. Thus, the state obtained by the application of (\ref{U}) to the seeded state $\ket{\psi_o}$ is given by

\begin{equation}\label{State}
\ket{\psi}= \hat U \hat D(\alpha) \ket{0}. 
\end{equation}
As we can see in Appendix A, this expression can be obtained using the asymptotic treatment of \cite{SipeVelho}, with no other assumption than the undepleted pump regime, which is justified when $|\chi|^2 \ll 1$ (recall that $\gamma = {\cal A}\chi)$. 

We can now rewrite the operators appearing in Eq. (\ref{State}) as $\hat D(\alpha) \hat D^{\dagger}(\alpha) \hat U \hat D(\alpha) \ket{0}$ and consider that the seed's mode is orthogonal to the signal's. As a result, the former does not affect the latter and: 

\begin{eqnarray}\label{Disp}
&&\hat D^{\dagger}(\alpha) \hat U \hat D(\alpha) = \tilde {\hat U}= \\
&&e^{\gamma \int \int L({\bf k}, {\bf k'})\hat a_s^{\dagger}({\bf k})\left [\hat a_i^{\dagger}({\bf k'})+\alpha({\bf k'})\right ]{\rm d}{\bf k} {\rm d}{\bf k'}-h.c.}\nonumber
\end{eqnarray}
If the signal mode is detected close to a given value ${\bf k'_s}$, with a measurement resolution $\delta {\bf k'_s}$, the intensity of the detected signal can be expressed as:

\begin{equation}\label{detect}
\langle \hat a_s^{\dagger}({\bf k'_s})\hat a_s({\bf k'_s})\rangle \delta {\bf k'_s}= \bra{0} \tilde {\hat U}^{\dagger} \hat a_s^{\dagger}({\bf k'_s})\hat a_s({\bf k'_s}) \tilde {\hat U} \ket{0}\delta {\bf k'_s},
\end{equation}
where we have used that $\left [\hat a ({\bf k_s}) ,\hat a^{\dagger} ({\bf k'_i}) \right ] = 0$ and that the displacement operator associated to the seed commutes with the signal modes. 

It is now convenient to express all the creation and annihilation operators using the Schmidt decomposition. We have then that 
$\hat S = \gamma \int \int L({\bf k},{\bf k'})\hat a^{\dagger}_s({\bf k}) \hat a^{\dagger}_i({\bf k'}){\rm d}{\bf k}{\rm d}{\bf k'}
=\gamma\sum_n \sqrt{\lambda_n}\hat b^{\dagger}_n\hat c^{\dagger}_n$, where $\hat b^{\dagger}_n = \int \psi_n({\bf k})\hat a^{\dagger}_s({\bf k}){\rm d}{\bf k}$ and $\hat c^{\dagger}_n = \int \phi_n({\bf k})\hat a^{\dagger}_i({\bf k}){\rm d}{\bf k}$ are the Schmidt modes (see Appendix B and \cite{PhysRevLett.84.5304}, for instance, for the full details of the change to the Schmidt basis).

Using the previous results and definitions, we can thus expand, in first place, the operator $\hat a({\bf k_s})$ in the Schmidt basis as follows :
\begin{equation}\label{SchmidtA}
\hat a_s({\bf k_s})=\sum_n \psi_n({\bf k_s}) \hat b_n.
\end{equation}
Equivalently, the seed mode $\hat s= \int \frac{\alpha^*({\bf k})}{|\alpha|^2}\hat a_i({\bf k}){\rm d}{\bf k}=\sum_n \alpha^*_n \hat c_n$ can also be expressed in this same basis for convenience, with  $\alpha_n = \int \frac{\alpha({\bf k})}{|\alpha|^2}\phi_n^*({\bf k}){\rm d}{\bf k} $ or, alternatively, $\alpha({\bf k})=|\alpha|^2 \sum_n \alpha_n \phi_n({\bf k})$.

Using the Schmidt decomposition, we can now express $\hat U^{\dagger}\hat a_s({\bf k_s}) \hat U$ as: 
\begin{equation}\label{SchmidtTransform}
\hat U^{\dagger}\hat a_s({\bf k_s}) \hat U=\sum_n \psi_n({\bf k_s}) \left[\hat b_n \cosh{(\gamma\sqrt{\lambda_n})}+\hat c_n^{\dagger}\sinh{(\gamma\sqrt{\lambda_n})}\right],
\end{equation}
and, consequently, 
\begin{eqnarray}\label{SchmidtTransformN}
&&\langle \hat U^{\dagger}\hat a_s^{\dagger}({\bf k_s}) \hat a_s({\bf k_s}) \hat U \rangle = \bra{0}\hat U^{\dagger}\hat a_s^{\dagger}({\bf k_s}) \hat a_s({\bf k_s}) \hat U\ket{0}= \nonumber \\
&&\sum_n |\psi_n({\bf k_s}) |^2\sinh^2{(\gamma\sqrt{\lambda_n})}.
\end{eqnarray}

We now consider the effect of the seed mode. It performs the transformation $\hat c_n \rightarrow \hat c_n + \alpha_n$, so that \eqref{SchmidtTransformN} finally becomes: 
\begin{eqnarray}\label{SchmidtDisplace}
\langle \hat U^{\dagger}\hat a_s^{\dagger}({\bf k_s})\hat a_s({\bf k_s}) \hat U \rangle 
&& = \sum_n |\psi_n({\bf k_s}) |^2\sinh^2{(\gamma\sqrt{\lambda_n})}  \nonumber \\
& & +\abs{\alpha}^2\left|\sum_{n}\alpha^*_n\psi_n({\bf k_s}) \sinh{(\gamma\sqrt{\lambda_n})} \right|^2, \nonumber
\end{eqnarray}
which is the output of the produced two-mode (signal and idler) field.

We can now discuss Eq. \eqref{SchmidtDisplace}. By considering, for instance, that no mode selection is performed and detection is made in the signal mode,  we have that the detected intensity can be expressed as $\langle \hat n_s \rangle = \bra{0} \hat n_s \ket{0} = \int \langle \hat U^{\dagger}\hat a_s({\bf k_s}) ^{\dagger}\hat a_s({\bf k_s}) \hat U \rangle {\rm d}{\bf k_s}$, which leads to
\begin{equation}\label{SignalN}
\langle \hat n_s \rangle = \sum_{n}\sinh^2{(\gamma\sqrt{\lambda_n})}(1+|\alpha_n|^2|\alpha|^2).
\end{equation}
For $|\alpha|^2 \gg 1$, the first term in the r.h.s of \eqref{SchmidtDisplace} and \eqref{SignalN}, which corresponds to the spontaneous emission term, can be neglected, and we have that $\langle \hat n_s \rangle \approx \sum_{n}\sinh^2{(\gamma\sqrt{\lambda_n})}|\alpha_n|^2|\alpha|^2$. A first consequence of this expression is that  by shaping the seed beam only to a Schmidt mode, using techniques similar to the ones proposed in \cite{PhysRevA.101.033838}, one can measure $\lambda_n$ and infer the noise  and entanglement properties of the whole system. 

We now consider, as in \cite{Sipe} and Eq. \eqref{detect}, that the detection is performed in a given mode ${\bf k_s}$ in the interval $\delta {\bf k_s}$ of the signal mode. 
Eq. \eqref{SchmidtDisplace} then becomes: 
\begin{equation}\label{SchmidtDisplace2}
\langle \hat U^{\dagger}\hat a_s^{\dagger}({\bf k_s})\hat a_s({\bf k_s}) \hat U \rangle = 
|\alpha|^2\left|\sum_{n}\alpha^*_n\psi_n({\bf k_s})\sinh{(\gamma\sqrt{\lambda_n})}\right|^2.
\end{equation}
Several interesting specific cases can be studied from the expression above. If $\gamma \sqrt{\lambda_n} \ll 1$, we have that $\sinh{(\gamma\sqrt{\lambda_n})} \approx \gamma\sqrt{\lambda_n}$. In addition to obtain an expression which is valid for an arbitrary high value of $\gamma$, the benefit of the exact expression \eqref{SchmidtDisplace2} is to give a precise physical insight on the conditions of validity of the usual $\gamma \ll 1$ approximation.
Indeed, we see, that it is actually the factor $\gamma\lambda_n$ that plays the role of the control parameter.
We thus can infer, that the usual $\gamma \ll 1$ approximation can remain valid even with a higher pump field intensity if its spectrum and the non-linear interaction are such that the output field is highly entangled. Indeed as the number of entangled modes increases, the number of non-zero values of the $\lambda_n$ parameter also increases, always keeping $\sum_n\lambda_n = 1$. In the present contribution, we just point out this aspect that will not be discussed in details. 

Now, using that $\sqrt{\lambda_n}\psi_n({\bf k})=\int L({\bf k},{\bf k'})\phi_n^*({\bf k'}){\rm d}{\bf k'}$ we have finally that Eq. \eqref{SchmidtDisplace2} becomes
\begin{eqnarray}\label{SchmidtDisplace2a}
 &&\abs{\gamma\sum_{n} \int \int L({\bf k_s},{\bf k'}) \alpha({\bf k''}) \phi_n({\bf k''})\phi^*_n({\bf k'}){\rm d}{\bf k'} {\rm d}{\bf k'}}^2=  \nonumber \\
 &&\abs{\gamma \int L({\bf k_s},{\bf k'}) \alpha^*({\bf k'}){\rm d}{\bf k'}}^2. 
\end{eqnarray}
If $\alpha({\bf k'})$ has a flat spectrum in the support of $ L({\bf k_s},{\bf k'})$, we have that Eq. \eqref{SchmidtDisplace2a} returns the marginal of  $L({\bf k_s},{\bf k'}) $ with respect to the idler mode. For $\alpha({\bf k'}) \approx \delta ({\bf k})|\alpha|^2$, so that $|\alpha_n| \approx |\phi_n^*({\bf k})|$, we obtain the original result from \cite{Sipe}: 
\begin{equation}\label{SchmidtDisplace2b}
\langle \hat a_s^{\dagger}({\bf k_s})\hat a_s({\bf k_s})\rangle \approx \gamma^2 |\alpha|^2  |L({\bf k_s},{\bf k})|^2.
\end{equation}
 
We recall that even if from this result one can obtain information about the absolute value of the mode function at each point ${\bf k_s}, {\bf k}$  it is not obvious to obtain information about the phase of the function $ L({\bf k_s},{\bf k})$ for any mode.
 
In order to solve this problem, we now use the output state Eq. (\ref{State}) as the inputs of an interferometer.

\section{Interferometric detection}\label{Inter}

In this section we use the results obtained in the previous section in a configuration which leads to the direct measurement of the full joint modal function $L({\bf k},{\bf k'})$ of signal and idler beams.  For such, we assume that the signal and the idler fields are spatially separated and serve as the inputs of an interferometer. Then, the two modes are combined in a balanced beam-splitter, and the intensities of the output fields in modes $A$ and $B$ are detected.  

The field operators in the input modes ${s}$ and ${i}$ are combined as follows: $\hat a_A({\bf k})=(\hat a_s({\bf k})+\hat a_i({\bf k}))/\sqrt{2}$ and $\hat a_B({\bf k})=(\hat a_s({\bf k})-\hat a_i({\bf k}))/\sqrt{2}$. The difference between the field intensities detected at each one of the two output ports of the interferometer can be expressed as: 
 
\begin{eqnarray}\label{interf}
&&\langle (\hat N_A - \hat N_B)\rangle = \int \langle (\hat a_A^{\dagger}({\bf k})\hat a_A({\bf k})-\hat a_B^{\dagger}({\bf k})\hat a_B({\bf k}))\rangle{\rm d}{\bf k} =  \nonumber \\
&&\int \langle (\hat a_s^{\dagger}({\bf k})\hat a_i({\bf k})+\hat a_i^{\dagger}({\bf k})\hat a_s({\bf k}))\rangle{\rm d}{\bf k}.
\end{eqnarray}
which is thus the measurement of the non-resolved modal cross-correlation of the signal and idler fields. 

We can now use the Schmidt decomposition of \eqref{H} and of the idler mode to compute the effect of the non-linear evolution operator on $\hat a_i({\bf k)}$: 

\begin{equation}\label{SchmidtDisplaceIdler}
\hat U^{\dagger}\hat a_i({\bf k}) \hat U = \sum_n \phi_n({\bf k})
\left[\hat c_n \cosh(\gamma \sqrt{\lambda_n})+\hat b_n^{\dagger}\sinh{(\gamma \sqrt{\lambda_n})}\right].
\end{equation}
Then, considering as previously the effect of the seed field, that displaces the idler mode, we have that Eq. \eqref{interf} becomes:
\begin{eqnarray}\label{interfS}
 &&\langle (\hat N_A - \hat N_B)\rangle = \\
&&|\alpha|^2{\rm Re}\left [\int  \sum_{n,m} \psi_n^*({\bf k})\phi_m({\bf k})\alpha_n \alpha_m \sinh{(2\gamma \sqrt{\lambda_n})}){\rm d}{\bf k} \right ].\nonumber 
\end{eqnarray}
It is now convenient to re-express Eq. \eqref{interfS} by transforming the seed mode back from the Schmidt decomposition. By doing so, we can for instance study the limit $\gamma \ll 1$, that can be written as:
\begin{eqnarray}\label{end}
 &&\langle (\hat N_A - \hat N_B)\rangle = \\
&& 2{\rm Re}\left [\int \int  \gamma L({\bf k},{\bf k'})\alpha^*({\bf k})\alpha^*({\bf k'}){\rm d}{\bf k} {\rm d}{\bf k'} \right ].\nonumber
\end{eqnarray}

A first comment on Eq. (\ref{end}) is that even though it relates to the real part of a function, one can easily access the imaginary part of this function by simply  changing the phase of the signal beam, which can be done with standard experimental techniques. 

We can see that Eq. (\ref{end}) does not provide full  information about the modal function $L({\bf k}, {\bf k'})$. Nevertheless, studying some particular cases can be helpful to gain some intuition, and find out  how to experimentally access this function.

To start with, let's suppose  that the seed field's modal distribution is much broader than the width of the function $L({\bf k}, {\bf k'})$, and that it's nearly flat in the region of the support of $L({\bf k}, {\bf k'})$. In this case, we can take the seed  out of the integral in Eq. (\ref{end}), which becomes: 

\begin{eqnarray}\label{endflat}
&&\langle (\hat N_A - \hat N_B)\rangle = 2{\rm Re}\left [\gamma{\mathcal{B}}^2\int \int L({\bf k}, {\bf k'}){\rm d}{\bf k} {\rm d}{\bf k'} \right ]= \nonumber \\
&& 2{\rm Re}\left [{\mathcal{B}}^2 \tilde L(0, 0) \right ],
\end{eqnarray}
where $\tilde L({\bf q}, {\bf q'})= \int \int L({\bf k}, {\bf k'})e^{i{\bf q}{\bf k}}e^{i{\bf q'}{\bf k'}}{\rm d}{\bf k} {\rm d}{\bf k'}$ is the Fourier transform of the function $L({\bf k}, {\bf k'})$ and ${\mathcal{B}}$  is the value of the amplitude of the seed which is taken as a constant for ${\bf k}$ and ${\bf k'}$ in the support of  $ L({\bf k}, {\bf k'})$.

\begin{figure*}
 \begin{center}
\includegraphics[width=1\textwidth]{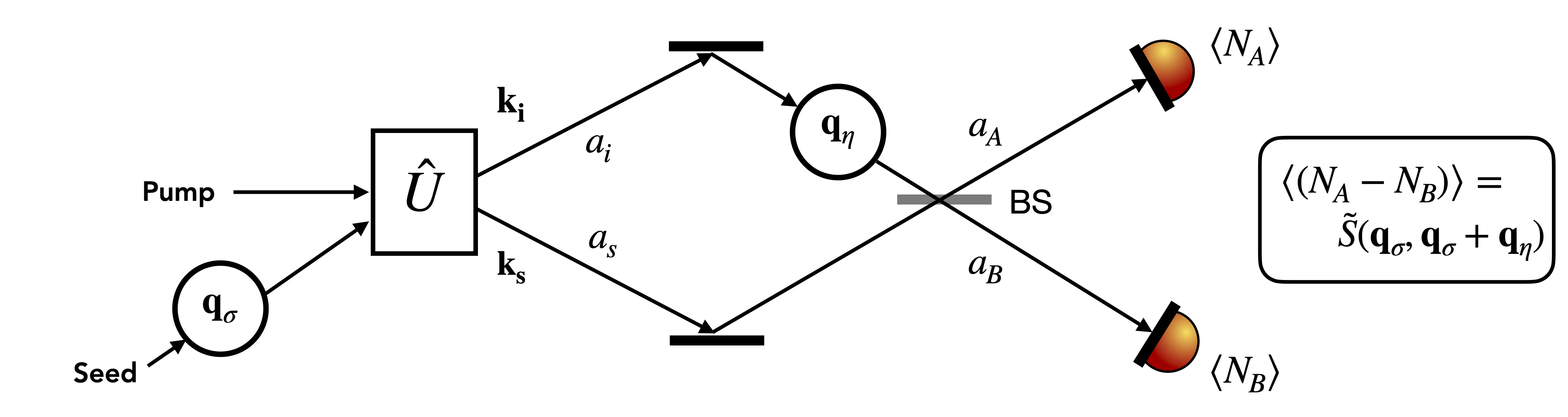}
\caption{\label{interfer}Interferometric detection for the joint temporal amplitude (JTA). In the time-frequency case, the parameters are $q_{\sigma}\equiv \tau$, $q_{\eta}\equiv t$ and BS stands for beam-splitter. A classical pump field and a coherent seed field generate a pair of modes noted $a_{i} $ and $a_{s}$. The seed is shifted of $q_{\sigma}$ before the stimulated emission process, and the output mode is then shifted from a value $q_{\eta}$. The two modes $a_{i} $ and $a_{s}$ are then recombined into a balanced beam-splitter, and non-resolved direct detections are performed in each final mode $a_{A}$ and $a_{B}$}
\end{center}
\end{figure*}

Eq. (\ref{endflat}) provides information about only one point of the modal distribution's Fourier transform, given by ${\bf q} = {\bf q'} = 0$. Of course this is not enough to completely characterize the modal state of the produced field. Accessing the whole modal function's Fourier transform requires measuring all the points of the function $\tilde L({\bf q}, {\bf q'})$. In order to do so, we will adapt the interferometer as depicted in Fig.~\ref{interfer}. 

We first apply a phase difference between the two arms of the interferometer and, say, the idler mode is transformed as $\hat a_i({\bf k}) \rightarrow \hat U^{\dagger}_{\bf q_{\eta}}\hat a_i({\bf k})\hat U_{\bf q_{\eta}}$, where $\hat U^{\dagger}_{\bf q_{\eta}}$ is the unitary operator that implements a phase difference between the two arms of the interferometer. This phase can be considered to be proportional to a parameter ${\bf q_{\eta}}$. By choosing $\hat U_{\bf q_{\eta}}$ such that $\hat U^{\dagger}_{\bf q_{\eta}}\hat a_i({\bf k})\hat U_{\bf q_{\eta}}= e^{{i\bf k}{\bf q_{\eta}}}\hat a_i({\bf k})$, it's easy to check that Eq. (\ref{end}) becomes: 

\begin{equation}\label{end2}
\langle (\hat N_A - \hat N_B)\rangle = 2{\rm Re}\left [\gamma\int \int L({\bf k}, {\bf k'})\alpha^*({\bf k})\alpha^*({\bf k'})e^{{i\bf k'}{\bf q_{\eta}}}{\rm d}{\bf k} {\rm d}{\bf k'} \right ]. 
\end{equation}

We can now add another phase factor, but this time to the seed pulse, and {\it before} it interacts with the non-linear medium.  This corresponds to providing a relative phase factor of ${\bf q_{\sigma}}$ between the seed and the pump beam, so that $\alpha({\bf k}) \rightarrow \alpha({\bf k})e^{-i{\bf k}{\bf q_{\sigma}}}$. The complete measurement scheme is  represented in Fig. \ref{interfer}, where all the phase factors are identified. It consists of a configuration analogous to the one implemented in \cite{Zela, Paulao}, where it is shown that transformations implemented to the seed's transverse field profile are transferred to the signal's one.

The produced output signal is then given by: 

\begin{eqnarray}\label{end3}
&&\langle (\hat N_A - \hat N_B)\rangle = \tilde S({\bf q_{\sigma}}, {\bf q_{\sigma} }+ {\bf q_{\eta}})= \\
&& 2{\rm Re}\left [\gamma\int \int L({\bf k}, {\bf k'})\alpha^*({\bf k})e^{i{\bf k}{\bf q_{\sigma}}}\alpha^*({\bf k'})e^{i{\bf k'} {\bf q_{\sigma}}}e^{{i\bf k'}{\bf q_{\eta}}}{\rm d}{\bf k} {\rm d}{\bf k'} \right ], \nonumber
\end{eqnarray}
where $\tilde S({\bf q_1}, {\bf q_2})$ is the Fourier transform of the function $L({\bf k_1}, {\bf k_2})\alpha^*({\bf k_1})\alpha^*({\bf k_2})$. As previously, we first discuss the situation of a seed beam with a broad and flat modal distribution, which leads directly to: 

\begin{eqnarray}\label{endFlat}
&&\langle (\hat N_A - \hat N_B)\rangle = \tilde L({\bf q_{\sigma}}, {\bf q_{\sigma}} + {\bf q_{\eta}})= \\
&&2{\rm Re}\left [\gamma{\cal B}^2\int \int L({\bf k}, {\bf k'})e^{i{\bf k} {\bf q_{\eta}}}e^{i{\bf k'} {\bf q_{\sigma}}}e^{{i\bf k'}{\bf q_{\sigma}}}{\rm d}{\bf k} {\rm d}{\bf k'} \right ]. 
\nonumber
\end{eqnarray}

The function $\tilde L({\bf q_{\sigma}}, {\bf q_{\sigma}} + {\bf q_{\eta}})$  is  the Fourier transform of the modal function  $ L({\bf k}, {\bf k'})$ at points ${\bf q_{\sigma}}$ and ${\bf q_{\sigma}} + {\bf q_{\eta}}$. As a consequence, once the real part of $\tilde L({\bf q_{\sigma}}, {\bf q_{\sigma}} + {\bf q_{\eta}})$ is also measured, the proposed scheme enables the complete measurement of the modal function $ L({\bf k}, {\bf k'})$  and/or its Fourier transform, including its phase information. We can see that the required modal properties of the seed are the opposite of the ones used in the direct detection configuration. While in that case the seed was supposed to have a narrow modal distribution, in the present one we considered it to have a  broad, nearly flat one. This is somehow intuitive, since in the direct detection configuration we accessed the absolute value of the function $ L({\bf k}, {\bf k'})$, while in the interferometric set-up we can measure its Fourier transform $\tilde L({\bf q_{\sigma}}, {\bf q_{\sigma}} + {\bf q_{\eta}})$. 

As a matter of fact, the flat modal distribution requirement can be loosened, since one can use the measured signal (\ref{end3}) at different points and invert it to obtain $ L({\bf k}, {\bf k'})$  (provided that the seed's spectrum is known, of course). Another option is to simply deconvolute the $\tilde \alpha({\bf q})$ functions (the Fourier transform of the $\alpha({\bf k})$ functions) from Eq. (\ref{end3})). Supposing that we have measured both real and imaginary parts of the signal, and defining 

\begin{eqnarray}
&&\tilde S({\bf q_{\sigma}},{\bf q_{\eta}}+ {\bf q_{\sigma}})= \\
&& 2 \gamma\int \int L({\bf k}, {\bf k'})\alpha^*({\bf k'})\alpha^*({\bf k})e^{i{\bf k} {\bf q_{\sigma}}}e^{i{\bf k'} {\bf q_{\eta}}}e^{{i\bf k'}{\bf q_{\sigma}}}{\rm d}{\bf k} {\rm d}{\bf k'}, \nonumber
\end{eqnarray}
we have that 

\begin{equation}\label{Invert}
L({\bf k}, {\bf k'})= \frac{1}{8 \pi^2 \gamma}\frac{ \int \int \tilde S({\bf q_{\sigma}}, {\bf q_{\sigma}}) e^{-i{\bf k} {\bf q_{\sigma}}}e^{-i{\bf k'} {\bf q_{\eta}}}e^{{-i\bf k'}{\bf q_{\sigma}}}{\rm d}{\bf k} {\rm d}{\bf k'} }{\alpha^*({\bf k'})\alpha^*({\bf k})},
\end{equation} 
thus providing a direct measurement of the modal function at all points. 

We have presented a general method to directly measure the function $L({\bf k}, {\bf k'})$ and its Fourier transform. The proposed solution provides all the modal properties associated to the field produced by a quadratic interaction between a non-linear medium and a pump beam in the low-gain regime using intensity measurements only, circumventing coincidence measurements. We notice that errors due to expansion on $\gamma$ depend on $\gamma^3$. A discussion on precision and sampling requirements and the effects of noise and measurement imperfections can be found in Appendix C.

The presented description and discussion apply to any degree of freedom of the electromagnetic field in the low-gain regime and in the undepleted pump approximation. In the next section, we will discuss in details the case of frequency and time degrees of freedom. 

\section{Joint temporal amplitude reconstruction}\label{JTASec}

We now study in details the case of frequency and time modes, which is particularly interesting due to the experimental difficulties in manipulating these degrees of freedom, and thus measuring the modal function in all points with phase information, when compared for instance to the polarization or the transverse position and momentum. Also, frequency modes and states have many applications in different experimental set-ups described by the interaction Eq.(\ref{U}). In the intense pump coupling regime, for instance, it is possible to create multimode squeezed states of the radiation  both in the bulk or in the circuit configurations \cite{roslund_wavelength-multiplexed_2014,  yang_squeezed_2021, PhysRevLett.107.030505,Virginia, yokoyama_ultra-large-scale_2013}, and the choice of different frequency mode basis adapt the squeezed and highly mode-entangled produced state  \cite{pinel_ultimate_2012, pinel_quantum_2013} to applications as quantum metrology, quantum information and quantum communication. 

In the weak coupling regime, as the SPDC one, frequency also plays an important role. In this case,  the unitary operator (\ref{U}) is expanded until the first order and the produced state is  post-selected by coincidence measurements, which excludes the vacuum part of the state. The resulting post-selected state  becomes a valuable non-gaussian resource in quantum optics and quantum information. This state can be, for instance, highly entangled in different modes, as polarization - that can be used to encode qubits - or frequency - that can be used to encode qubits, qudits  \cite{Qudits, PhysRevX.5.041017}, or for continuous variables for quantum computing \cite{PhysRevA.102.012607, fabre:tel-03191301, fabre_time-frequency_2022} and  metrology \cite{chen_hong-ou-mandel_2019}.

In order to analyze the time and frequency modes case, we change notations so that  $ L({\bf k}, {\bf k'}) \equiv  L(\omega, \omega')$ and $\tilde L({\bf q_{\eta}}, {\bf q_{\sigma}}) \equiv \tilde L(\tau, t )$ are, respectively,  the JSA and the Joint Temporal Amplitude (JTA). ln order to do so, we now identify how the different terms appearing in Eq. (\ref{end3}) can be experimentally implemented. The first point to be addressed is the pump's spectral width. By inverting Eq. (\ref{end3}) and using this Section's notation, we see that 

\begin{equation}\label{JTA}
L(\omega, \omega')= \frac{1}{8 \pi^2 \gamma}\frac{ \int \int \tilde S(\tau, t + \tau) e^{-i \tau \omega} e^{-i (t +\tau) \omega'}{\rm d} t {\rm d} \tau }{\alpha^*(\omega)\alpha^*(\omega')}
\end{equation} 
can be inferred by varying $t$ and $\tau$ and reconstructing $\tilde S(\tau, t + \tau)$ from the measurement results. Of course, $\alpha(\omega)$ should be known, and it must have a support larger than the one of $L(\omega, \omega')$, otherwise  $\tilde S(\tau, t + \tau)$ cannot be reconstructed.

The second point to be discussed is how to implement the temporal delay $\tau$. This term is associated to an optical path difference between the pump and the seed (see Fig. \ref{interfer}). We can for instance assume that both pump and seed come from the same laser and the pump is frequency doubled while the seed is attenuated. In this case, we ensure the phase coherence between both classical beams, and the possibility to choose the relative phase proportional to the parameter $\tau$ between both wave-packets. This time delay between pump and seed beams can occur {\it before} the non-linear interaction, and is inherited by the photons created by it. This is a situation similar to the one studied in \cite{Zela} but in the context of the beam's transverse profile. 

Finally, experimentally, the phase difference proportional to the time delay $t$ can be obtained {\it after} the non-linear interaction by placing an optical path difference in the idler's arm, as is usually done in interferometers.

Of course, one cannot determine $t$ and $\tau$ with infinite precision. A discussion of the experimental requirements can be found in Appendix C for arbitrary modes, together with the main idea of how to sample results experimentally. 

As a final comment, the proposed method can also, in principle, be used to directly infer the JSA from the JTA simply by reversing the roles of frequency and time in the discussion above.

\section{Discussion and conclusion}

We have extended the idea of SET to arbitrary intensity regimes, avoiding approximations other than the undepleted pump one. From our results, we have proposed a method to obtain full phase information about the modal function of the field produced by a non-linear quadratic interaction using an interferometric configuration. The obtained results are  analogous to a two dimensional homodyne detection but acting on the mode profile of the field instead of its quadratures. We have discussed in detail the case of the time and spectral properties of the field as well as the role of noise and imperfections (see Appendix C). 

We believe that the present contribution will be useful to different experimental configurations and set-ups where the modal properties of the field play a role for quantum information, metrology and imaging.

\bibliography{biblioSET2}
\onecolumngrid
\appendix

\section{Field quantum state generated by the nonlinear device}\label{detail}
In this appendix we show that the quantum state of the light $\ket{\psi_{\text{out}}}$ emitted by the non linear device
can be written as
\beq
\ket{\psi_{\text{out}}} = \hat{U} \ket{\psi_{\text{in}}} = e^{\hat{S} - \text{h.c.}} \ket{\psi_{\text{in}}},
\label{eq:outUin}
\eeq
where $\ket{\psi_{\text{in}}}$ is the quantum state of the light incident upon the non linear device and where
\[
\hat{S}  =  \gamma \sum_{\nu,\eta}\int d\vec{k}_1 d\vec{k}_2 L_{\nu,\eta}(\vec{k}_1,\vec{k}_2) \hat{b}^+_{\nu \vec{k}_1}  \hat{b}^+_{\eta \vec{k}_2}.
\]
In this last relation, the couple $(\nu,\vec{k})$ labels a plane wave mode of the field, with wave-vector $\vec{k}$ and $\nu=\pm 1$ labels the two orthogonal polarization. The function $L_{\nu,\eta}(\vec{k}_1,\vec{k}_2)$ which characterizes the state of the emitted field is such that
$\sum_{\nu, \eta}\int d\vec{k}_1 d\vec{k}_2 \abs{L_{\nu,\eta}(\vec{k}_1,\vec{k}_2)}^2 = 1$. The factor $\gamma $ can be written as $\gamma=\chi\mathcal{A}$, where $\chi$ is a characteristic of the nonlinear device and  $\abs{\chi}^2$ represents the number of photon pair generation per photon of the pump pulse. $\mathcal{A}$ is a characteristic of the pump pulse and   $\abs{\mathcal{A}}^2$ is the mean number of photons in the pump pulse. Therefore $\abs{\gamma}^2$ is the total number of photon pairs generated by the pump pulse.

As in Ref.~\cite{Yang2008} or Ref.~\cite{Sipe}, we consider that the device is characterized by a Hamiltonian  $\hat{H}_{\text{L}} + \hat{H}_{\text{NL}}$, where $\hat{H}_{\text{L}}$ collects the linear interactions responsible for the propagation and dispersion and $\hat{H}_{\text{NL}}$ collecting the non linear interactions is given by
\beq
\hat{H}_{\text{NL}} = -\left[\sum_{\nu,\eta,\mu}\int d\vec{k}_1 d\vec{k}_2 d\vec{k} S_{\nu \eta \mu}(\vec{k}_1,\vec{k}_2,\vec{k}) \hat{b}^+_{\nu \vec{k}_1}  \hat{b}^+_{\eta \vec{k}_2}  \hat{a}_{\mu \vec{k}} + \text{h.c.} \right],
\label{eq:HNL}
\eeq
where the function $S_{\nu \eta \mu}(\vec{k}_1,\vec{k}_2,\vec{k})$ encompasses the non linear coupling and phase matching condition.
We proceed as in Ref.~\cite{Yang2008}, where only the spontaneous process is considered and we generalize to the stimulated case, where the incident field is not necessarily the vacuum. For the sake of  self-contentedness we briefly summarize the methodology which relies on scattering theory and backward propagation.
The state $\ket{\psi_{\text{in}}}$ and $\ket{\psi_{\text{out}}}$ are scattering states~\cite{Yang2008, liscidini2012}. 
$\ket{\psi_{\text{in}}}$ is the field state at $t=0$ which develop from $\ket{\psi(t_-)}$
but considering that the nonlinear device has been removed. That is,
$\ket{\psi_{\text{in}}} = \exp(-i\hat{H}_{\text{L}}(0-t_-)) \ket{\psi(t_-)}$ (throughout this appendix, we consider $\hbar=1$).
In the same way, $\ket{\psi_{\text{out}}}$ is the state at $t=0$ which will develop to $\ket{\psi(t_+)}$, without the nonlinear device. That is~:
$\ket{\psi(t_+)} = \exp(-i\hat{H}_{\text{L}}(t_+-0))\ket{\psi_{\text{out}}}$. The times $t_-$ and $t_+$ are considered as $t_{\pm} \rightarrow \pm \infty$.
In that way, the effect of the nonlinear device is completely described by the transition
from $\ket{\psi_{\text{in}}}$ to $\ket{\psi_{\text{out}}}$ at $t=0$.

The relation between the "in" and the "out" states can be written as~:
\beq
\label{psioutUpsiin}
\ket{\psi_{\text{out}}} = \lim_{t_{\pm}\rightarrow \pm\infty}\hat{U}(t_+,t_-)\ket{\psi_{\text{in}}}
\eeq
where $\hat{U}(t_+,t_-)$ is given by
\beq
\hat{U}(t_+,t_-) = \exp(i\hat{H}_{\text{L}}t_+) \exp(-i\hat{H}(t_+-t_-))\exp(-i\hat{H}_{\text{L}}t_-)
\label{eq:Udef}
\eeq
The unitary operator $\hat{U}(t_+,t)$, as a function of $t$, fulfills the following differential equation,
\beq
\label{eq:dU}
-i \frac{\partial}{\partial t} \hat{U}(t_+,t)=  \hat{U}(t_+,t)\hat{V}(t) 
\eeq
with 
\[
\hat{V}(t) = \exp(i\hat{H}_{\text{L}}t) \hat{H}_{\text{NL}} \exp(-i\hat{H}_{\text{L}}t)
\]
Using Eq.~\eqref{eq:HNL}, we have
\beq
V(t) = -\left[\sum_{\nu,\eta,\mu}\int d\vec{k}_1 d\vec{k}_2 d\vec{k} S_{\nu \eta \mu}(\vec{k}_1,\vec{k}_2,\vec{k})\exp\left[-i(\omega_{\mu\vec{k}} - \omega_{\nu\vec{k}_1} - \omega_{\eta\vec{k}_2} )t\right] \hat{b}^+_{\nu \vec{k}_1}  \hat{b}^+_{\eta \vec{k}_2}  \hat{a}_{\mu \vec{k}} + \text{h.c.} \right].
\label{eq:V}
\eeq
As in ref.~\cite{Yang2008}, we will use back-propagation (from $t_+$ to $t_-$) because our objective is to express  expectation value of observables which are written as a function of output modes  annihilation and creation operators (at $t_+ \rightarrow +\infty$).

Before starting the calculations, let state the assumptions.
We suppose that the function $S_{\nu \eta \mu}(\vec{k}_1,\vec{k}_2,\vec{k})$ characterizing the non linear device in Eq.~\eqref{eq:HNL}, is non zero only when pump  $(\mu,\vec{k})$, the signal   $(\nu,\vec{k}_1)$ and the idler $(\eta,\vec{k}_2)$ modes satisfy the phase matching condition imposed by the nonlinear medium and the device geometry.
Furthermore, the following assumptions are in order~:
\begin{itemize}
	\item[A1] The pump mode is orthogonal to the other modes, seed, signal and idler. Moreover, the frequency range of the pump mode  does not overlap with other modes.
	\item[A2] The seed mode overlaps with the idler mode  and is orthogonal to the signal mode. 
	\item [A3] The probability to emit a photon pair, per pump photon, is very small.
\end{itemize}


To obtain $\ket{\psi_{\text{out}}}$
using Eq.~\eqref{psioutUpsiin}, we must specify  the state "in" state of the field $\ket{\psi_{\text{in}}}$. 
We consider that before entering in the nonlinear device the state of the field is well described by a coherent state, corresponding to the pump and the seed pulse.
Because of assumptions A1 and A2 we can write the "in" state  as
$\ket{\psi_{\text{in}}} = \ket{\psi_p}\otimes\ket{\psi_s}$, where $\ket{\psi_p}$ refers to the pump pulse coherent state and $\ket{\psi_s}$ to the seed coherent pulse.
To describe the field states corresponding to the pump pulse, we first define the creation operator $A[f_p]^+$ as~:
\beq
\label{eq:modeCreation}
\hat{A}^+[f_p] = \sum_\mu \int f_p(\mu,\vec{k}) \hat{a}^+_{\mu,\vec{k}}d\vec{k}
\eeq
where the function $f_p(\mu,\vec{k})$ characterizes the mode of the pump  and is normalized as $\sum_{\mu}\int\abs{f_p(\mu,\vec{k})}^2d\vec{k}=1$.
The "in" field state  $\ket{\psi_p}$ corresponding to the pump pulse is then written as $\ket{\psi_p} = D[\mathcal{A},f_p]\ket{\text{vac}}$, where the displacement operator $D[\mathcal{A},f_p]$ is defined as:
\[
D[\mathcal{A},f_p] = 	\exp\left(\mathcal{A}A^+[f_p] - \text{h.c.}\right)
\]
and $\abs{\mathcal{A}}^2$ is the mean number of photon in the mode $f_p$, that is $\bra{\psi_p}\sum_{\mu}\int a^+_{\mu\vec{k}}a_{\mu\vec{k}}d\vec{k}\ket{\psi_p} =\abs{\mathcal{A}}^2$.

The seed pulse coherent state $\ket{\psi_s}$ is defined in the same way but with the help of the normalized mode function $f_s(\mu, \vec{k})$, and the corresponding displacement operator $D[\mathcal{B},f_s]$ as $\ket{\psi_s} = D[\mathcal{B},f_s] \ket{\text{vac}}$,
where~:
\[
D[\mathcal{B},f_s] = 	\exp\left(\mathcal{B}B^+[f_s] - \text{h.c.}\right).
\]
The creation  operator $B^+[f_s]$ is defined as in Eq.~\eqref{eq:modeCreation} but replacing $f_p$ by $f_s$~:
\beq
\label{eq:modeCreation}
B^+[f_s] = \sum_\mu \int f_s(\mu,\vec{k}) b^+_{\mu,\vec{k}}d\vec{k}
\eeq
We have used the notation $b^+_{\mu,\vec{k}}$ for the creation operators to remember that 
they commute with the annihilation operator $a_{\mu,\vec{k}}$ because the support of the function $f_s$ and $f_p$ are orthogonal, $\sum_\mu\int d\vec{k} f_s^*(\mu,\vec{k})f_p(\mu,\vec{k}) = 0$,
by assumption~A1.

Finally, the "in" state is thus written as:
\beq
\ket{\psi_{\text{in}}} = D[\mathcal{A},f_p]\otimes D[\mathcal{B},f_s]\ket{\text{vac}}.
\label{eq:psiin}
\eeq
Setting $\mathcal{B}=0$ gives the spontaneous parametric conversion process and when $\mathcal{B}\neq 0$ the stimulated process is considered.

To calculate the "out" state we follow the work by Yang et al. in~\cite{Yang2008}.
Let's  define two operators:   
\beq
K_a= \mathcal{A}A^+[f_p]; \quad K_b=  \mathcal{B} B^+[f_s]
\label{eq:Kdef}
\eeq
then 
\begin{align}
\ket{\psi_{\text{out}}} = U(t_+,t_-)e^{K_a-h.c.}e^{K_b-h.c.}\ket{\text{vac}} &= 
U(t_+,t_-)e^{K_a-h.c.}e^{K_b-h.c.}U^{\dagger}(t_+,t_-)\ket{\text{vac}} \nonumber \\
= 
e^{\overline{K}_a(t_-)- h.c.}e^{\overline{K}_b(t_-)-h.c.}\ket{\text{vac}}
\label{eq:psiout}
\end{align}
where 
\beq
\hat{\overline{K}}_{a(b)}(t) = \hat{U}(t_+,t) \hat{K}_{a(b)}\hat{U}^{\dagger}(t_+,t)
\label{eq:backprop}
\eeq
and the second equality in Eq.~\eqref{eq:psiout} follows from the fact that $H_{\text{L}}\ket{\text{vac}} = H_{\text{NL}}\ket{\text{vac}} = 0$.

The back-propagated operator $\hat{\overline{K}}_{a(b)}(t)$ satisfies~:
\beq
i\dt{\hat{\overline{K}}_{a(b)}(t)} = [\hat{\overline{K}}_{a(b)}(t),\hat{\overline{V}}(t)] 
\label{eq:derBackprop}
\eeq
where $\hat{\overline{V}}(t) = \hat{U}(t_+,t) \hat{V}(t)\hat{U}^{\dagger}(t_+,t)$. Using Eq.~\eqref{eq:Udef}, Eq.~\eqref{eq:HNL} and~Eq.~\eqref{eq:V},  it can be written as
\beq
\hat{\overline{V}}(t) = -\left[\sum_{\nu,\eta,\mu}\int d\vec{k}_1 d\vec{k}_2 d\vec{k} S_{\nu \eta \mu}(\vec{k}_1,\vec{k}_2,\vec{k})\exp\left[-i(\omega_{\mu\vec{k}} - \omega_{\nu\vec{k}_1} - \omega_{\eta\vec{k}_2} )t\right] \hat{\overline{b}}^+_{\nu \vec{k}_1}(t)  \hat{\overline{b}}^+_{\eta \vec{k}_2}(t)  \hat{\overline{a}}_{\mu \vec{k}}(t) + \text{h.c.} \right]
\label{eq:Vbar}
\eeq
where the bar above an operator means its back-propagation as in Eq.~\eqref{eq:backprop}.
In the following, it is convenient to distinguish the pump and seed modes annihilation operators  as  $c_{\mu, \vec{k}}$ and $d_{\mu,\vec{k}}$ respectively. In the following calculation the commutators of these operators can be considered as zero, that is $[c_{\mu, \vec{k}},d_{\eta,\vec{k}'}] = [c_{\mu, \vec{k}}^+,d_{\eta,\vec{k}'}]=0$, because the ranges of $\vec{k}$ and $\vec{k}'$ will never overlap. With this notation $\overline{V}(t)$ can be rewritten as
\begin{align}
\hat{\overline{V}}(t) &= -\left[\sum_{\nu,\eta,\mu}\int d\vec{k}_1 d\vec{k}_2 d\vec{k} S_{\nu \eta \mu}(\vec{k}_1,\vec{k}_2,\vec{k})\exp\left[-i(\omega_{\mu\vec{k}} - \omega_{\nu\vec{k}_1} - \omega_{\eta\vec{k}_2} )t\right] \hat{\overline{b}}^+_{\nu \vec{k}_1}(t)  \hat{\overline{b}}^+_{\eta \vec{k}_2}(t)  \hat{\overline{c}}_{\mu \vec{k}}(t) + \text{h.c.} \right]
\label{eq:Vbarcd} \nonumber \\
&- \left[\sum_{\nu,\eta,\mu}\int d\vec{k}_1 d\vec{k}_2 d\vec{k} S_{\nu \eta \mu}(\vec{k}_1,\vec{k}_2,\vec{k})\exp\left[-i(\omega_{\mu\vec{k}} - \omega_{\nu\vec{k}_1} - \omega_{\eta\vec{k}_2} )t\right] \overline{b}^+_{\nu \vec{k}_1}(t)  \hat{\overline{b}}^+_{\eta \vec{k}_2}(t)  \hat{\overline{d}}_{\mu \vec{k}}(t) + \text{h.c.} \right]
\end{align}
The back-propagated creation and annihilation operators all fulfill the same differential equation as the one fulfilled by $\overline{K}(t)$ given in Eq.~\eqref{eq:derBackprop}. In addition, in the integral we have the following commutation relations (see rules A1 and A2)
\begin{align}
&[\hat{\overline{c}}_{\mu \vec{k}}(t),\hat{\overline{b}}^+_{\nu \vec{k}_1}(t)]=
[\hat{\overline{c}}^+_{\mu \vec{k}}(t),\hat{\overline{b}}^+_{\nu \vec{k}_1}(t)]=
[\hat{\overline{c}}_{\mu \vec{k}}(t),\hat{\overline{b}}^+_{\eta \vec{k}_2}(t)]=
[\hat{\overline{c}}^+_{\mu \vec{k}}(t),\hat{\overline{b}}^+_{\eta \vec{k}_2}(t)]=
0 \nonumber \\
&[\hat{\overline{d}}_{\mu \vec{k}}(t),\hat{\overline{b}}^+_{\nu \vec{k}_1}(t)]=
[\hat{\overline{d}}^+_{\mu \vec{k}}(t),\hat{\overline{b}}^+_{\nu \vec{k}_1}(t)]=0
\end{align}
The overlap of the seed mode with the idler mode can be characterized by an overlap function $f_{\mu,\mu'}(\vec{k},\vec{k}')$ as:
\beq
[\hat{\overline{d}}^+_{\mu \vec{k}}(t),\hat{\overline{b}}_{\mu' \vec{k}'}(t)] = f_{\mu,\mu'}(\vec{k},\vec{k}')
\eeq
Therefore, the differential equations fulfilled by $\hat{\overline{c}}^+_{\mu \vec{k}}(t)$, $\hat{\overline{d}}^+_{\mu \vec{k}}(t)$ and $\hat{\overline{b}}^+_{\mu \vec{k}}(t)$ are
\begin{align}
\label{eq:diffeqsforc}
i \dt{}\hat{\overline{c}}^+_{\mu \vec{k}}(t) &= 
\sum_{\nu,\eta}\int d\vec{k}_1 d\vec{k}_2  S_{\nu \eta \mu}(\vec{k}_1,\vec{k}_2,\vec{k})\exp\left[-i(\omega_{\mu\vec{k}} - \omega_{\nu\vec{k}_1} - \omega_{\eta\vec{k}_2} )t\right] \hat{\overline{b}}^+_{\nu \vec{k}_1}(t)  \hat{\overline{b}}^+_{\eta \vec{k}_2}(t)   \\
i\dt{}\overline{d}^+_{\mu \vec{k}}(t) &= 
\sum_{\nu,\eta,\mu'}\int d\vec{k}_1  d\vec{k}_2d\vec{k}' f_{\mu,\eta}(\vec{k},\vec{k}_2) S^*_{\nu \eta \mu'}(\vec{k}_1,\vec{k}_2,\vec{k}')\exp\left[i(\omega_{\mu'\vec{k}'} - \omega_{\nu\vec{k}_1} - \omega_{\eta\vec{k}_2} )t\right] 
\hat{\overline{c}}^+_{\mu' \vec{k}'}(t)
\hat{\overline{b}}_{\nu \vec{k}_1}(t)  \nonumber \\
&+
\sum_{\nu,\eta,\mu'}\int d\vec{k}_1 dd\vec{k}_2\vec{k}' f_{\mu,\eta}(\vec{k},\vec{k}_2) S^*_{\nu \eta \mu'}(\vec{k}_1,\vec{k},\vec{k}')\exp\left[i(\omega_{\mu'\vec{k}'} - \omega_{\nu\vec{k}_1} - \omega_{\eta\vec{k}_2} )t\right] 
\hat{\overline{d}}^+_{\mu' \vec{k}'}(t)
\hat{\overline{b}}_{\nu \vec{k}_1}(t) \nonumber \\
&+
\sum_{\nu,\eta}\int d\vec{k}_1 d\vec{k}_2  S_{\nu \eta \mu}(\vec{k}_1,\vec{k}_2,\vec{k})\exp\left[-i(\omega_{\mu\vec{k}} - \omega_{\nu\vec{k}_1} - \omega_{\eta\vec{k}_2} )t\right] \hat{\overline{b}}^+_{\nu \vec{k}_1}(t)  \hat{\overline{b}}^+_{\eta \vec{k}_2}(t)
\label{eq:diffeqsford}  \\
i\dt{}\hat{\overline{b}}^+_{\nu \vec{k}_1}(t) &=  
\sum_{\eta,\mu}\int d\vec{k} d\vec{k}_2  
\left(
S^*_{\nu \eta \mu}(\vec{k}_1,\vec{k}_2,\vec{k}) +
S^*_{\eta \nu \mu}(\vec{k}_2,\vec{k}_1,\vec{k}) 
\right)
\exp\left[i(\omega_{\mu\vec{k}} - \omega_{\nu\vec{k}_1} - \omega_{\eta\vec{k}_2} )t\right]
\left(\hat{\overline{c}}^+_{\mu \vec{k}}(t)  + \hat{\overline{d}}^+_{\mu \vec{k}}(t)\right)  \hat{\overline{b}}_{\eta \vec{k}_2}(t) 
\end{align}
In principle, solving these differential equations with the initial conditions $\hat{\overline{d}}^+_{\mu \vec{k}}(t_+) = \hat{d}^+_{\mu \vec{k}}$, 
$\hat{\overline{c}}^+_{\mu \vec{k}}(t_+) = \hat{c}^+_{\mu \vec{k}}$ and $\hat{\overline{b}}^+_{\mu \vec{k}}(t_+) = \hat{b}^+_{\mu \vec{k}}$
allows to compute $\hat{\overline{K}}_{a(b)}(t_-)$ defined by Eqs.~\eqref{eq:backprop} and \eqref{eq:Kdef} and then $\ket{\psi_{\text{out}}}$ by Eq.~\eqref{eq:psiout}.
Indeed, 
\beq
\hat{\overline{K}}_{a}(t_-) = \mathcal{A}\sum_\mu \int d\vec{k} f_p(\mu,\vec{k}) \hat{\overline{c}}^+_{\mu \vec{k}}(t_-); \quad 
\hat{\overline{K}}_{b}(t_-) = \mathcal{B}  \sum_\mu \int d\vec{k} f_s(\mu,\vec{k}) \hat{\overline{d}}^+_{\mu \vec{k}}(t_-)
\label{eq:Kabbar}
\eeq
Using assumption [A3], as in Ref.~\cite{Yang2008} we solve the differential  Eqs.~\eqref{eq:diffeqsforc}~\eqref{eq:diffeqsford} at first order in $V$ with time dependent perturbation theory. We obtain
\begin{align}
\label{eq:cfirstOrder}
\hat{\overline{c}}^+_{\mu \vec{k}}(t_-) &\simeq \hat{c}^+_{\mu \vec{k}} +
\frac{1}{i}\sum_{\nu,\eta}\int d\vec{k}_1 d\vec{k}_2  S_{\nu \eta \mu}(\vec{k}_1,\vec{k}_2,\vec{k})\left[\int_{t_+}^{t_-}\exp\left[-i(\omega_{\mu\vec{k}} - \omega_{\nu\vec{k}_1} - \omega_{\eta\vec{k}_2} )t\right] dt\right] b^+_{\nu \vec{k}_1}  b^+_{\eta \vec{k}_2}   \\
\hat{\overline{d}}^+_{\eta \vec{k}}(t_-) &\simeq 
\hat{d}^+_{\mu \vec{k}}
+\frac{1}{i}
\sum_{\nu,\eta,\mu'}\int d\vec{k}_1  d\vec{k}_2d\vec{k}' f_{\mu,\eta}(\vec{k},\vec{k}_2) S^*_{\nu \eta \mu'}(\vec{k}_1,\vec{k}_2,\vec{k}')
\left[\int_{t_+}^{t_-}
\exp\left[i(\omega_{\mu'\vec{k}'} - \omega_{\nu\vec{k}_1} - \omega_{\eta\vec{k}_2} )t\right] dt \right]
\hat{c}^+_{\mu' \vec{k}'}
\hat{b}_{\nu \vec{k}_1}  \nonumber \\
&+\frac{1}{i}
\sum_{\nu,\eta,\mu'}\int d\vec{k}_1  d\vec{k}_2d\vec{k}' f_{\mu,\eta}(\vec{k},\vec{k}_2) S^*_{\nu \eta \mu'}(\vec{k}_1,\vec{k}_2,\vec{k}')
\left[\int_{t_+}^{t_-} \exp\left[i(\omega_{\mu'\vec{k}'} - \omega_{\nu\vec{k}_1} - \omega_{\eta\vec{k}_2} )t\right] dt \right]
\hat{d}^+_{\mu' \vec{k}'}
\hat{b}_{\nu \vec{k}_1}  \nonumber \\
&+\frac{1}{i}
\sum_{\nu,\eta}\int d\vec{k}_1 d\vec{k}_2  S_{\nu \eta \mu}(\vec{k}_1,\vec{k}_2,\vec{k})\left[\int_{t_+}^{t_-} \exp\left[-i(\omega_{\mu\vec{k}} - \omega_{\nu\vec{k}_1} - \omega_{\eta\vec{k}_2} )t\right] dt\right]
\hat{b}^+_{\nu \vec{k}_1}  \hat{b}^+_{\eta \vec{k}_2} 
\label{eq:dfirstOrder}
\end{align}
where all the operators appearing on the  right side of the equality are taken at time $t=t_+$.

We note that the operator $\overline{d}^+_{\eta \vec{k}}(t_-)$ will be applied to the vacuum, therefore the second and third term of Eq.~\eqref{eq:dfirstOrder} don't give any contribution.

The last term of Eq.~\eqref{eq:dfirstOrder} will give a non zero contribution only for frequencies $\omega_{\mu\vec{k}}$ in the range of the pump frequencies (this is a property of function $S_{\nu \eta \mu}(\vec{k}_1,\vec{k}_2,\vec{k})$).
But at the end, the operator $\overline{d}^+_{\eta \vec{k}}(t_-)$ will be used in the integral of Eq.~\eqref{eq:Kabbar} defining $\overline{K}_{b}(t_-)$, where the effective range of integration  is limited by the support of the function $f_s(\mu, k)$ defining the seed mode. Because of hypothesis A1 this range does not overlap with the one of the pump. The resulting integral is therefore zero.
Therefore, we obtain that
$\overline{d}^+_{\eta \vec{k}}(t_-) = d^+_{\mu \vec{k}}$ and
\beq
\hat{c}^+_{\mu \vec{k}}(t_-) \simeq \hat{c}^+_{\mu \vec{k}} +
\frac{2\pi}{i}\sum_{\nu,\eta}\int d\vec{k}_1 d\vec{k}_2 
S_{\nu \eta \mu}(\vec{k}_1,\vec{k}_2,\vec{k})\delta(\omega_{\mu\vec{k}} - \omega_{\nu\vec{k}_1} - \omega_{\eta\vec{k}_2} )
\hat{b}^+_{\nu \vec{k}_1}  \hat{b}^+_{\eta \vec{k}_2}  
\eeq

Finally, we obtain the expected result for $\ket{\psi_{\text{out}}}$~:
\begin{align}
\ket{\psi_{\text{out}}} &= \exp\left(\mathcal{A}\sum_\mu \int d\vec{k} f_p(\mu,\vec{k}) \hat{c}^+_{\mu \vec{k}}- \text{h.c..}\right)\ket{\text{vac}}\otimes \nonumber \\
&\exp\left[\left(\mathcal{A}\chi \sum_{\nu,\eta}\int d\vec{k}_1 d\vec{k}_2 L_{\nu,\eta}(\vec{k}_1,\vec{k}_2) \hat{b}^+_{\nu \vec{k}_1}  \hat{b}^+_{\eta \vec{k}_2} - \text{h.c..} \right)\right]
\exp\left(\mathcal{B}  \sum_\mu \int d\vec{k} f_s(\mu,\vec{k}) \hat{d}^+_{\mu \vec{k}} - \text{h.c..} \right)\ket{\text{vac}}
\end{align}
where we have defined the normalized function  $L_{\nu,\eta}(\vec{k}_1,\vec{k}_2)$ as
\begin{equation}\label{finalappendixspectrum}
L_{\nu,\eta}(\vec{k}_1,\vec{k}_2) = 
-\frac{2\pi i}{\chi} \sum_{\mu}\int d\vec{k} 
f_p(\mu,\vec{k})S_{\nu \eta \mu}(\vec{k}_1,\vec{k}_2,\vec{k})\delta(\omega_{\mu\vec{k}} - \omega_{\nu\vec{k}_1} - \omega_{\eta\vec{k}_2} )
\end{equation}
with $\chi$ a normalization factor defined as
\beq
\chi = 2\pi\left(\sum_{\nu, \eta}\int d\vec{k}_1 d\vec{k}_2
\abs{\sum_{\mu}\int d\vec{k} 
	f_p(\mu,\vec{k})S_{\nu \eta \mu}(\vec{k}_1,\vec{k}_2,\vec{k})\delta(\omega_{\mu\vec{k}} - \omega_{\nu\vec{k}_1} - \omega_{\eta\vec{k}_2} )}^2\right)^{\frac{1}{2}},
\eeq
such that
\[
\sum_{\nu, \eta}\int d\vec{k}_1 d\vec{k}_2 \abs{L_{\nu,\eta}(\vec{k}_1,\vec{k}_2)}^2 = 1.
\]
$\abs{\chi}^2$  is the probability of a photon pair generation per pump pulse photon in mode $f_p(\mu,\vec{k})$. Thus, $\abs{\chi \mathcal{A}}^2$ is the total number of generated pairs of photon.
We thus can identify the coefficient $\gamma$ in main text (see also Eq.~\eqref{eq:outUin})  to the product $\chi\mathcal{A}$.
We stress out that the validity of perturbation expansion (Eq.~\eqref{eq:cfirstOrder} and Eq.~\eqref{eq:dfirstOrder}) is independent of the intensities $\abs{\mathcal{A}}^2$ and $\abs{\mathcal{B}}^2$ of the pump and the seed respectively. It relies only on the smallness of $\abs{\chi}^2$ (assumption [A3]).

\section{Schmidt decomposition}\label{AppSchmidt}

We provide here some useful information on the Schmidt mode decomposition. In order to perform a basis change so as to decompose \eqref{H} in the Schmidt basis we used that $\hat S = \gamma \int \int L({\bf k},{\bf k'})\hat a^{\dagger}_s({\bf k}) \hat a^{\dagger}_i({\bf k'}){\rm d}{\bf k}{\rm d}{\bf k'}=\sum_n \sqrt{\lambda_n}\hat b^{\dagger}_n\hat c^{\dagger}_n$, where $\hat b^{\dagger}_n = \int \psi_n({\bf k})\hat a^{\dagger}_s({\bf k}){\rm d}{\bf k}$ and $\hat c^{\dagger}_n = \int \phi_n({\bf k})\hat a^{\dagger}_i({\bf k}){\rm d}{\bf k}$ are the Schmidt modes.

In addition, $L({\bf k},{\bf k'}) = \sum_n \sqrt{\lambda_n}\psi_n({\bf k})\phi_n({\bf k'})$. $\psi_n$ and $\phi_n$ form a basis in the spaces of the signal and the idler modes, respectively, with $\sqrt{\lambda_n}= \int \int L({\bf k},{\bf k'})\psi^*_n({\bf k})\phi^*_n({\bf k'}){\rm d}{\bf k}{\rm d}{\bf k'}$ and $\sqrt{\lambda_n}\psi_n({\bf k})=\int L({\bf k},{\bf k'})\phi^*_n({\bf k'}){\rm d}{\bf k'}$ and analogously for $\phi_n({\bf k})$. 

Using these relations, one can reproduce the results presented in Sections \ref{Direct} and \ref{Inter} of this manuscript.

\section{Noise}\label{Noise}

We now discuss possible measurement imperfections. For this, we'll consider two independent sources of imperfections : the first one can be seen as noise in the ${\bf q_i}$ variables, and may have different physical origins depending on the considered set-up. For example, when measuring the spectral properties of photons, ${\bf q} \equiv t $, which is an optical path delay that is centered around some value, say, $\tau$ with a distribution $P(\tau+\delta \tau)$, where $\delta \tau$ is distributed around $\tau$. The second one concerns the intrinsic discreteness of physical apparatuses, which means that the parameter ${\bf q}$ varies in finite steps, and ${\bf q_{i+1}}= {\bf q_{i}}+\Delta{\bf q}$. Taking again as an example the case of spectral measurement, this would correspond to the minimum path difference between the interfering pulses. 

We'll start by discussing the first case, that can be modeled as a random phase in the argument of, say, \eqref{endFlat}, as follows:

\begin{eqnarray}\label{Noise}
\int \int \int \int L({\bf k}, {\bf k'})e^{i{\bf k} ({\bf q_{\eta}}+\delta {\bf q_{\eta}})}e^{i{\bf k'} ({\bf q_{\eta}}+\delta {\bf q_{\eta}})}e^{{i\bf k'}({\bf q_{\sigma}}+\delta {\bf q_{\sigma}})}P({\delta {\bf q_{\eta}}})P({\delta {\bf q_{\sigma}}}){\rm d}{\bf k} {\rm d}{\bf k'}{\rm d}\delta {\bf q_{\sigma}}{\rm d}\delta {\bf q_{\eta}},
\end{eqnarray}
where $P({\delta {\bf q_{\alpha}}})$ is a distribution. We can for instance consider it as a Gaussian function, with $P({\delta {\bf q_{\alpha}}}) = {\cal N}e^{-\frac{\delta {\bf q_{\alpha}}^2}{\Delta_{\alpha}}}$. In this case, Eq. \eqref{Noise} can be easily computed, leading to: 

\begin{eqnarray}\label{Noise2}
\int \int  L({\bf k}, {\bf k'})e^{i{\bf k} {\bf q_{\eta}}}e^{i{\bf k'} {\bf q_{\eta}}}e^{{i\bf k'}{\bf q_{\sigma}}}e^{- ({\bf k}+{\bf k'})^2 \Delta_{\eta}}e^{- {\bf k'}^2 \Delta_{\sigma}}{\rm d}{\bf k} {\rm d}{\bf k'}.
\end{eqnarray}

The expression above sets a relationship between the frequency range that can be detected and the precision of the measurement, since the widths $\Delta^{-1}_{\eta}$ and $\Delta^{-1}_{\sigma}$ are the cut-off frequencies, setting an effective width to the integral \eqref{Noise2}. This is the usual condition observed in interferometric detection and can be met in different set-ups. 

We now discuss the effect of a finite variation of ${\bf q_{\alpha}}$ variables. This lead to a sampling of $ \tilde L({\bf q}, {\bf q'})$, in a number of (relevant) values that is proportional to the ratio between the width of the function  $ \tilde L({\bf q}, {\bf q'})$ and the intervals $\Delta{\bf q}$. A sampling rate of $1/(\Delta{\bf q})$ ensures the reconstruction of  $ \tilde L({\bf q}, {\bf q'})$ from its Fourier coefficients through the Nyquist–Shannon sampling theorem.

With these relations in mind, it is possible to estimate the requirements a given experimental set-up should meet to employ the presented method. 
\end{document}